\documentclass[finallayout,an,fleqn]{w-art}
\usepackage{times}
\usepackage{w-thm}
\theoremstyle{plain}

\theoremstyle{definition}

\usepackage[]{graphicx}
\chardef\bslash=`\\ 

\def\Granat{\hbox{\it GRANAT }}

\def\ergs{erg s$^{-1}$}

\begin{document}
\DOIsuffix{theDOIsuffix}
\Volume{324}
\Issue{S1}
\Copyrightissue{S1}
\Month{01}
\Year{2003}
\pagespan{1}{}
\Receiveddate{}
\Reviseddate{}
\Accepteddate{}
\Dateposted{}



\title[Review of LMXB near GC]{Review of low-mass X-ray binaries near the
  Galactic center}


\author[A. Lutovinov]{A. Lutovinov\footnote{e-mail: {\sf
      aal@hea.iki.rssi.ru}, Phone: +7\,095\,333\,2222, Fax:
    +7\,095\,333\,5377}\inst{1}} \address[\inst{1}]{Space Research
  Institute, 117997 Moscow, Profsoyuznaya str. 84/32, Russia}
\author[S. Grebenev]{S. Grebenev \inst{1}}
\author[S. Molkov]{S. Molkov\inst{1}}
\author[R. Sunyaev]{R. Sunyaev\inst{1,2}} \address[\inst{2}]{Max-Plank 
Institute for Astrophysics, Garching, Germany}
\begin{abstract}
Results of observations of several LMXBs in the Galactic center region
carried out with the ART-P telescope on board \Granat\ observatory are
briefly reviewed. More than dozen sources were revealed in this region
during five series of observations which were performed with the ART-P
telescope in 1990-1992. The investigation of the spectral evolution of
persistent emission of two X-ray bursters GX3+1 and KS1731-260, discussion
of QPO and spectral variations detected from the very bright Z-source GX5-1
and studying the pulse profile changes of the pulsar GX1+4 were carried out.

\end{abstract}
\maketitle                   





\section{Introduction}

The Galactic Center is one of the most interesting and intensively observing
in X-ray region populated by many sources, including dozens of low-mass
X-ray binaries. The ability of the telescope ART-P on board observatory
\Granat\ to investigate simultaneously X-ray emission from several point
like sources located within its field of view predetermined the choice of
the Galactic center as a favorable target for observations. The total
ART-P/\Granat\ exposure time of the Galactic center field observations was
$\sim$830 ks. Such a long exposure allowed to build the detailed X-ray map
of this region, to investigate persistent and transient sources emission, to
discover four new sources (Pavlinsky et al. 1994). Several years later this
region was observed by the telescopes of the BeppoSAX and RXTE observatories
(Sidoly et al. 1999, in't Zand 2001) in approximately the same energy band.
Now the INTEGRAL observatory begin to intensively investigate the Galactic
center in gamma rays.

In this paper we present briefly results of observations of two X-ray
bursters GX3+1 and KS1731-260 carried out with the ART-P/\Granat\ telescope
in 1990-1992. These objects are regular sources of type I bursts, which are
thought to be resulted from thermonuclear flares on the surface of accreting
neutron stars. We present here also the preliminary results of spectral and
timing investigations of very bright low-mass X-ray binary GX5-1 and X-ray
pulsar GX1+4.

\begin{vchfigure}[tb]
\includegraphics[width=10cm, height=10cm]{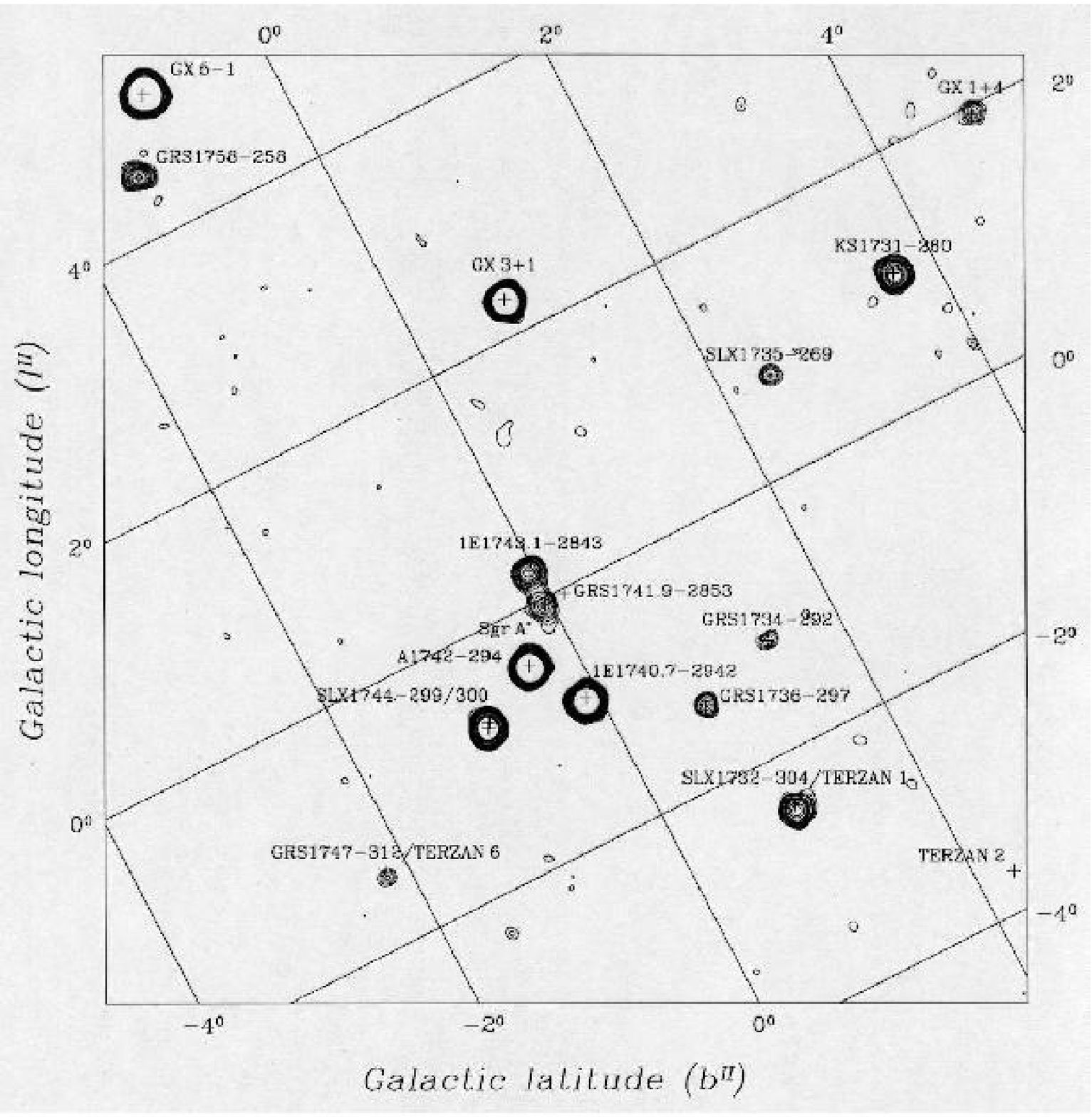}
\caption{Galactic center image obtained in the 3-20 keV energy band with the
  telescope ART-P in 1990. Contours show the significance of source
  detection at the level 3,4,5,10.... and more standard deviations. } 
\label{LutovinovA_gc}
\end{vchfigure}

\section{Results}

The image of the Galactic center field reconstructed on the combination of
the ART-P observations carried out in the fall of 1990 is presented in
Fig.\ref{LutovinovA_gc}.  Fifteen point sources were revealed. Most of them
were studied in several previous ART-P/\Granat\ papers (see for example
Pavlinsky et al. 1992, 1994, Grebenev et al. 1995, 1996, Lutovinov et
al. 2001, Molkov et al. 2000, 2001 and references there). Below we focus of
our attention on the four least studied sources.

\subsection{GX3+1}

This source was observed with ART-P four times during the \Granat\ Galactic
center field survey in the fall of 1990 with total exposure more than 60
ksec. The analysis of the data led to one interesting finding -- a strong
X-ray burst was detected from GX3+1 on 14 Oct. The investigation of the
source persistent emission in this day shows that its luminosity was $\sim
30$\% less than the luminosity measured in the other days and equal
$\sim5\times10^{37}\ \mbox{erg s}^{-1}$ in the 3-20 keV energy band. Such
behavior is in agreement with results of Makishima et al. (1983). It was
only second case after the {\it HAKUCHO} observations when GX3+1 was found
in the bursting state. Now the number of X-ray bursts detected from GX3+1
increased to 81; a part of these bursts were observed during source high
state (den Hartog et al. 2003).

\begin{figure}[tb]
\begin{minipage}[t]{.45\textwidth}
\includegraphics[bb=100 325 460 680, width=\textwidth]{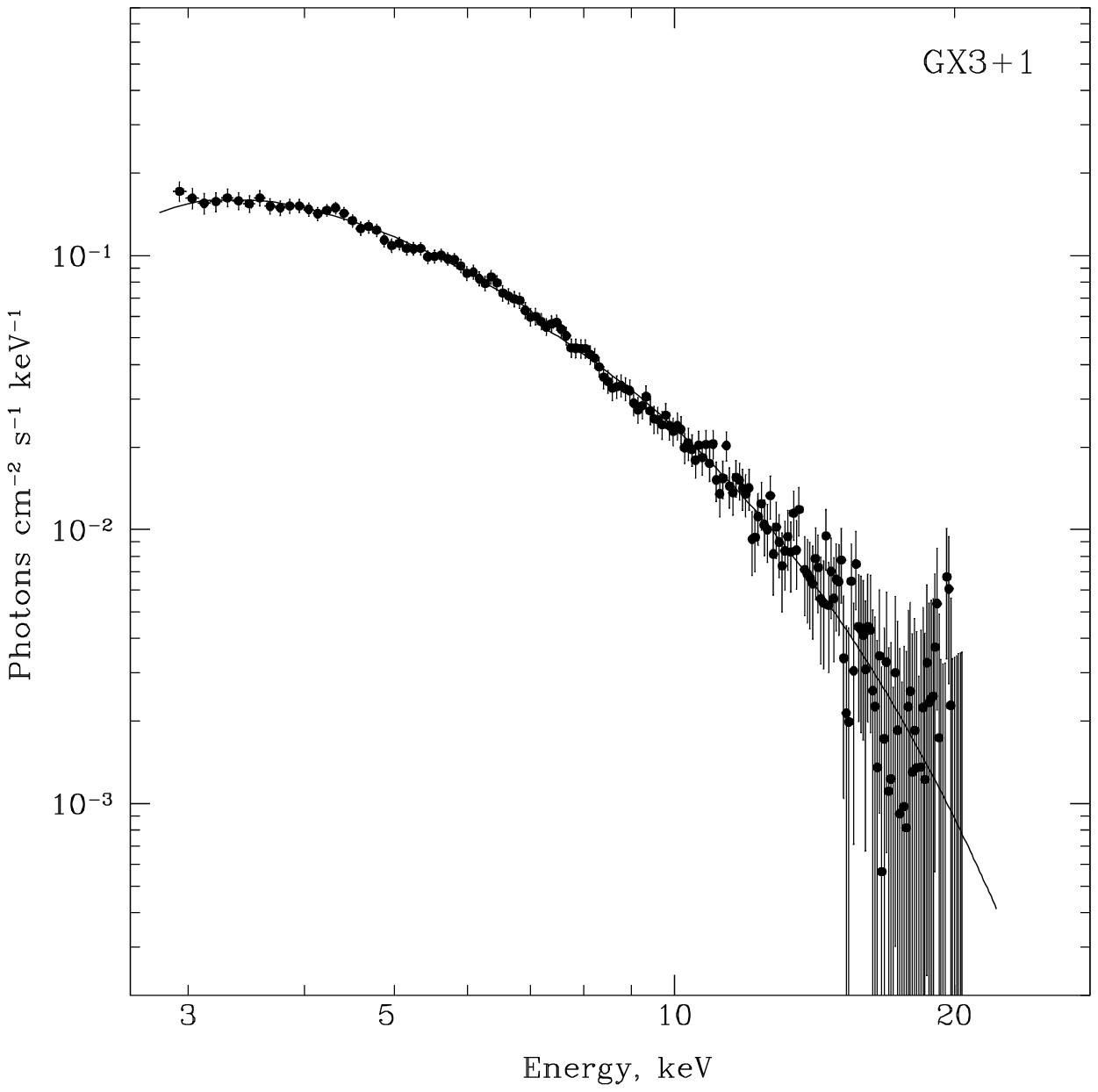}
\caption{Spectrum of GX3+1 measured by ART-P on Oct 11, 1990. Solid line is the
  best-fit comptonization model with temperature $kT=2.3$ keV. }
\label{LutovinovA_gx3_sp}
\end{minipage}
\hfil
\begin{minipage}[t]{.45\textwidth}
\includegraphics[bb=100 325 460 680, width=\textwidth]{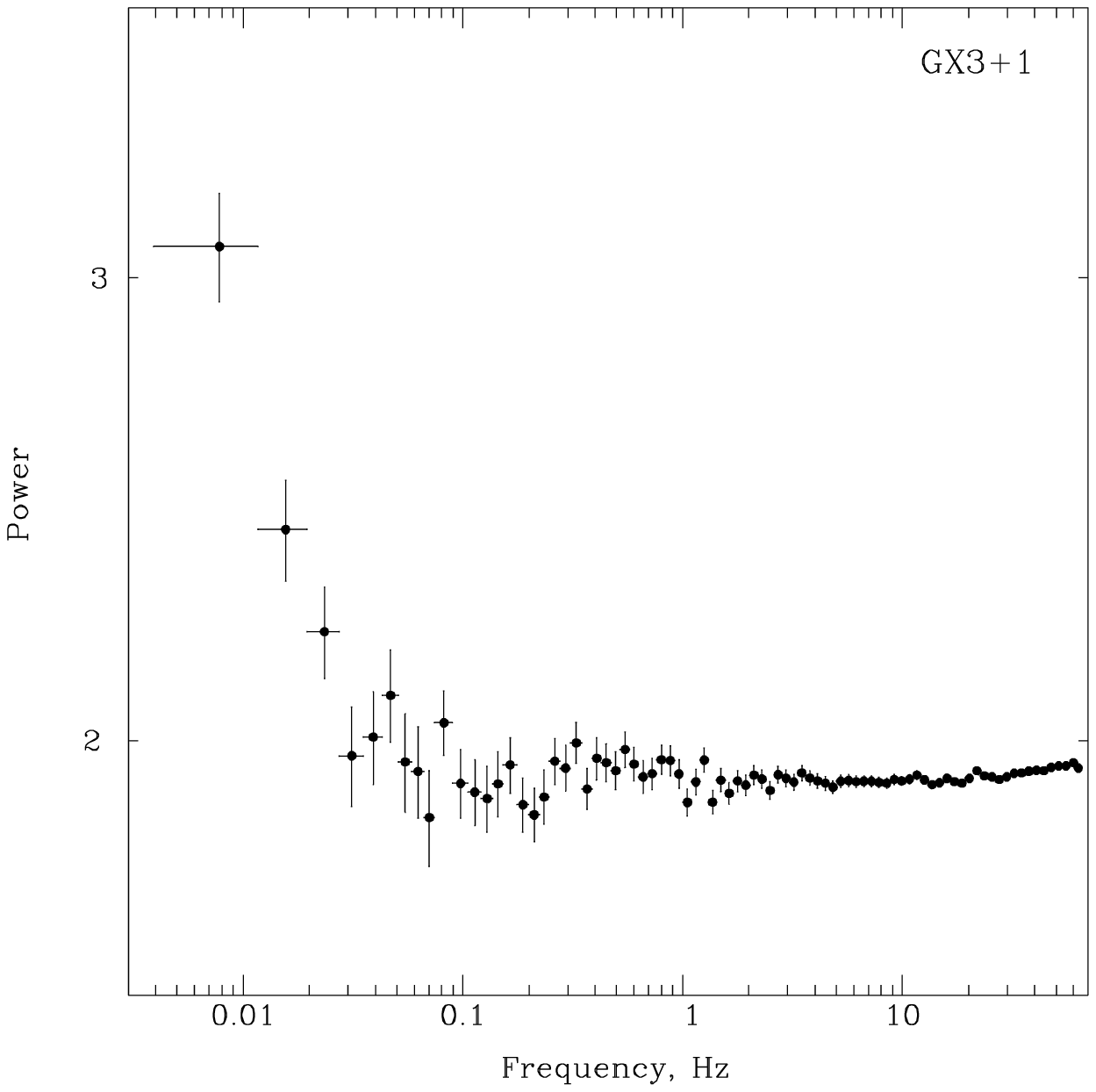}
\caption{Power spectrum (without dead-time correction) of the ART-P
  light curve of GX3+1 obtained on Oct 14, 1990.}
\label{LutovinovA_gx3_psp}
\end{minipage}
\end{figure}

The spectrum of the source measured in the high state is shown in
Fig.\ref{LutovinovA_gx3_sp}. It can be good approximated by the model of
comptonization of soft photons on hot electrons of plasma. However obtaining
values of parameters indicate that the model may be not physically correct
because it does not take into account effects of free-free absorption. Two
more suitable models for description of emission from GX3+1 were discussed
in Molkov et al. (1999). The power spectrum of GX3+1 is presented in
Fig.\ref{LutovinovA_gx3_psp} and doesn't demonstrate any peculiarities.

\subsection{KS1731-260}

The transient X-ray source KS1731-260 was discovered in 1989 with the
telescope TTM on board {\it ROENTGEN} observatory and was recognized as a
burster (Sunyaev et al. 1990). The source was observed many times with {\it
RXTE} and {\it BeppoSAX} observatories and coherent X-ray oscillations with
period $\sim1.9$ ms were discovered during several bursts (Smith et
al. 1997). Also a very powerful superburst with duration of several hours
was detected from the source (Kuulkers et al. 2002).
 
\begin{figure}[b]
\begin{minipage}[t]{.45\textwidth}
\includegraphics[bb=30 400 295 690, width=\textwidth]{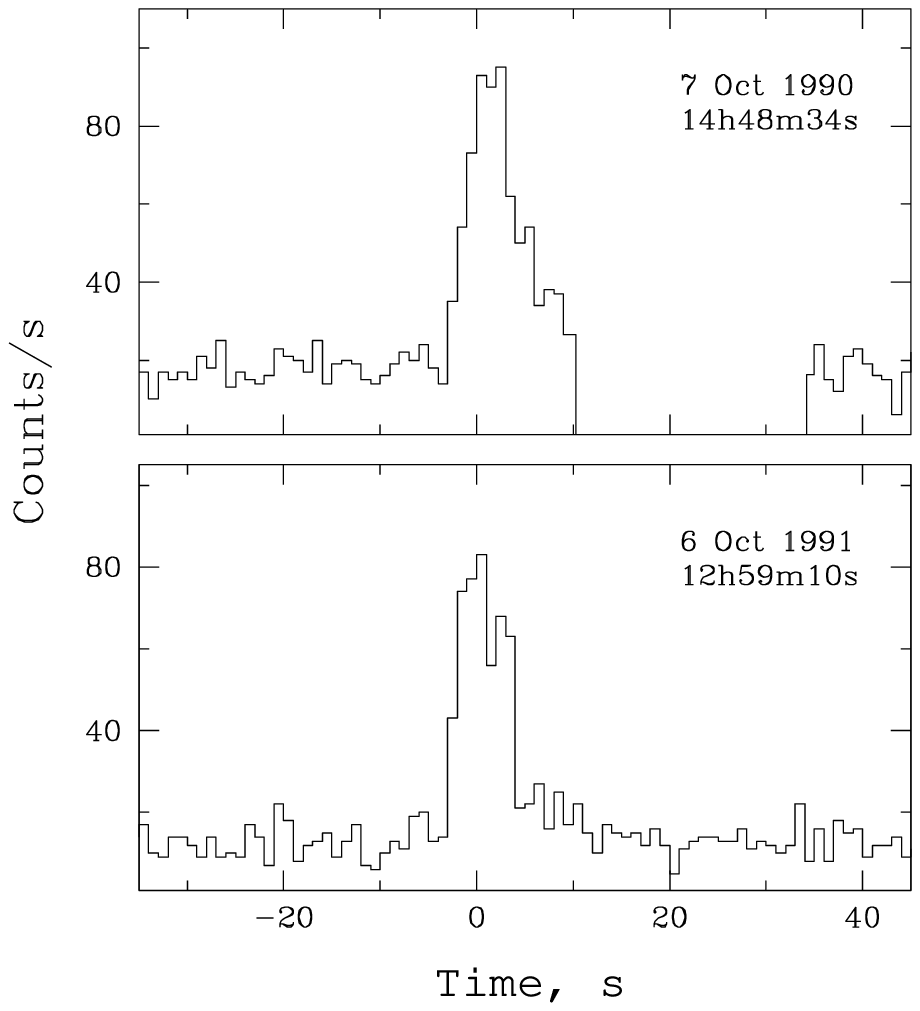}
\caption{Two X-ray bursts detected by ART-P from KS1731-260. Gap in the
  light curve is related to the data transfer from the telescope buffer to
  main satellite memory. }
\label{LutovinovA_k17_bur}
\end{minipage}
\hfil
\begin{minipage}[t]{.45\textwidth}
\includegraphics[bb=90 280 460 685, width=\textwidth]{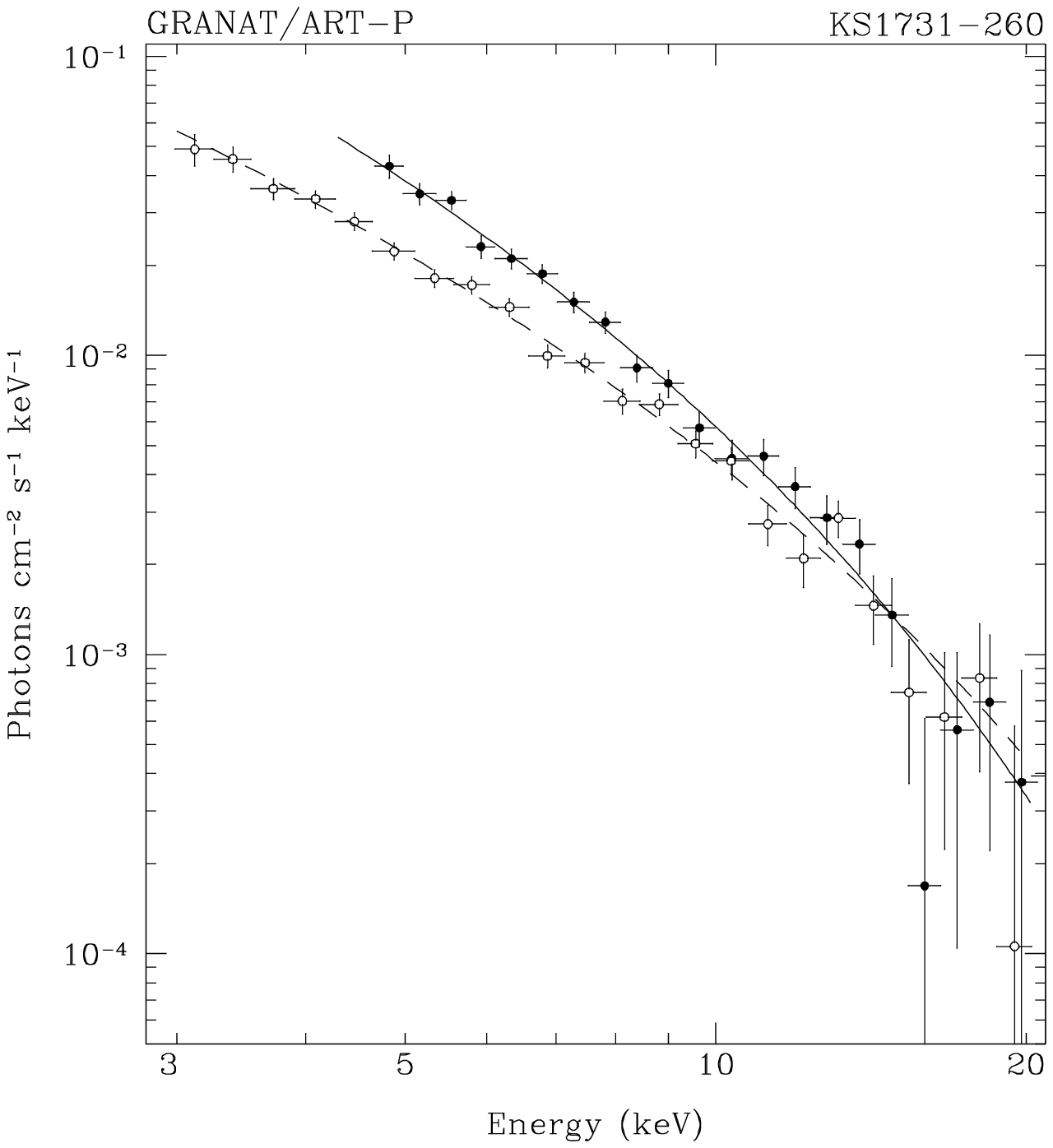}
\caption{Spectra of source persistent emission obtained on 23 Aug (open
  circles)  and 7 Oct (dark circles) 1990. The best-fit models are presented
  by corresponding lines.}
\label{LutovinovA_k17_sp}
\end{minipage}
\end{figure}

KS1731-260 was observed with the ART-P telescope four times with an interval
of one year. Two X-ray bursts with approximately the same time profile were
detected from the source (Fig.\ref{LutovinovA_k17_bur}). The photon spectra
of the source obtained on 23 Aug and 7 Oct 1990 are shown in
Fig.\ref{LutovinovA_k17_sp}.  In both cases the spectra were equally well
described by the Sunyaev-Titarchuk comptonization or optically thin thermal
bremsstrahlung models, but the source luminosity in the 4-20 keV energy band
measured on 7 Oct (when the X-ray burst was detected) was 30\% lower that
ones in another day. Such correlation between source low state and bursting
activity was observed for several other bursters.

The detailed analysis of the KS1731-260 persistent and bursts emissions is 
in progress.

\begin{vchfigure}[t]
\includegraphics[bb=115 275 415 685, width=.4\textwidth]{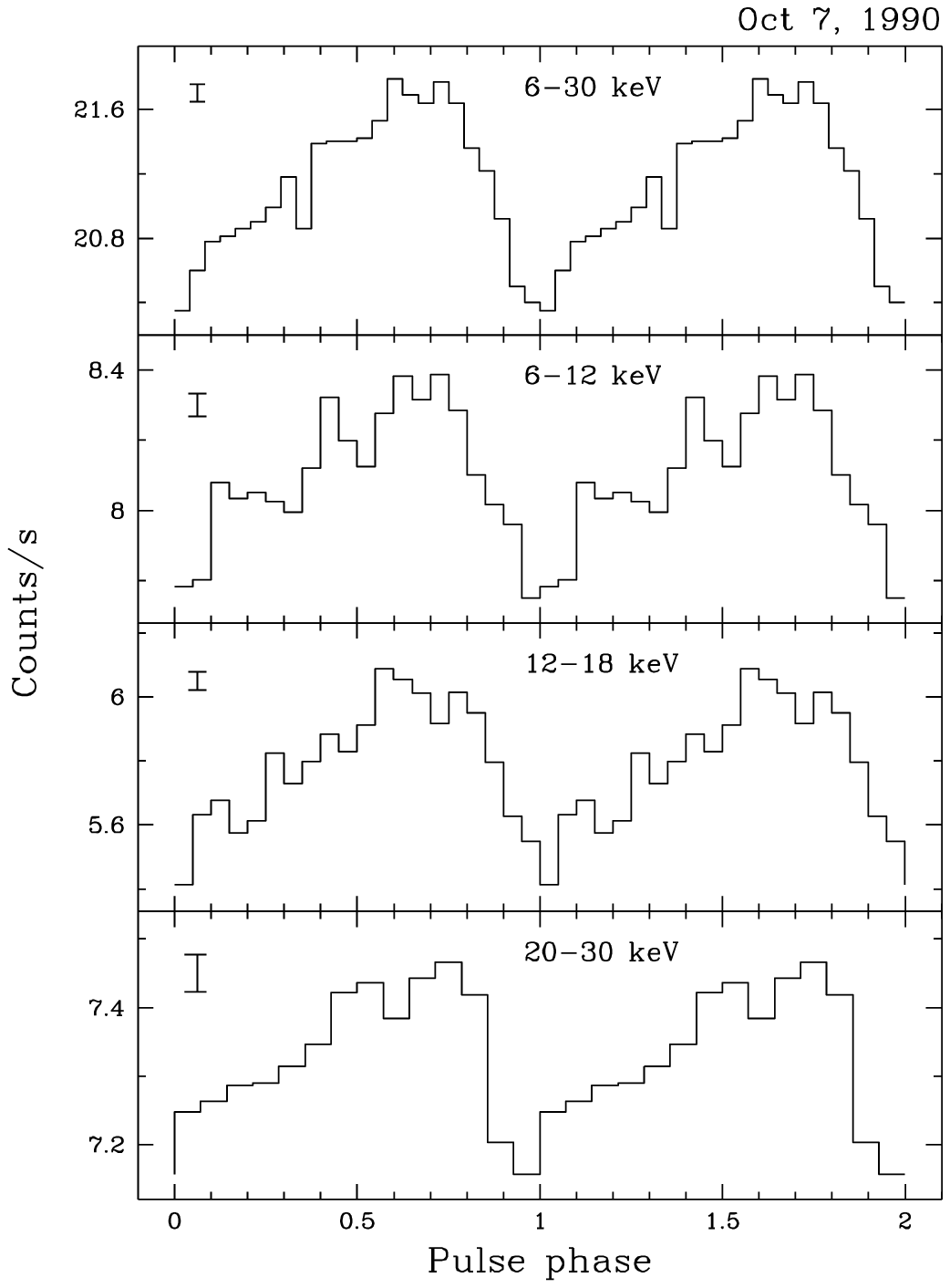}~a)
\hfil
\includegraphics[bb=115 275 415 685, width=.4\textwidth]{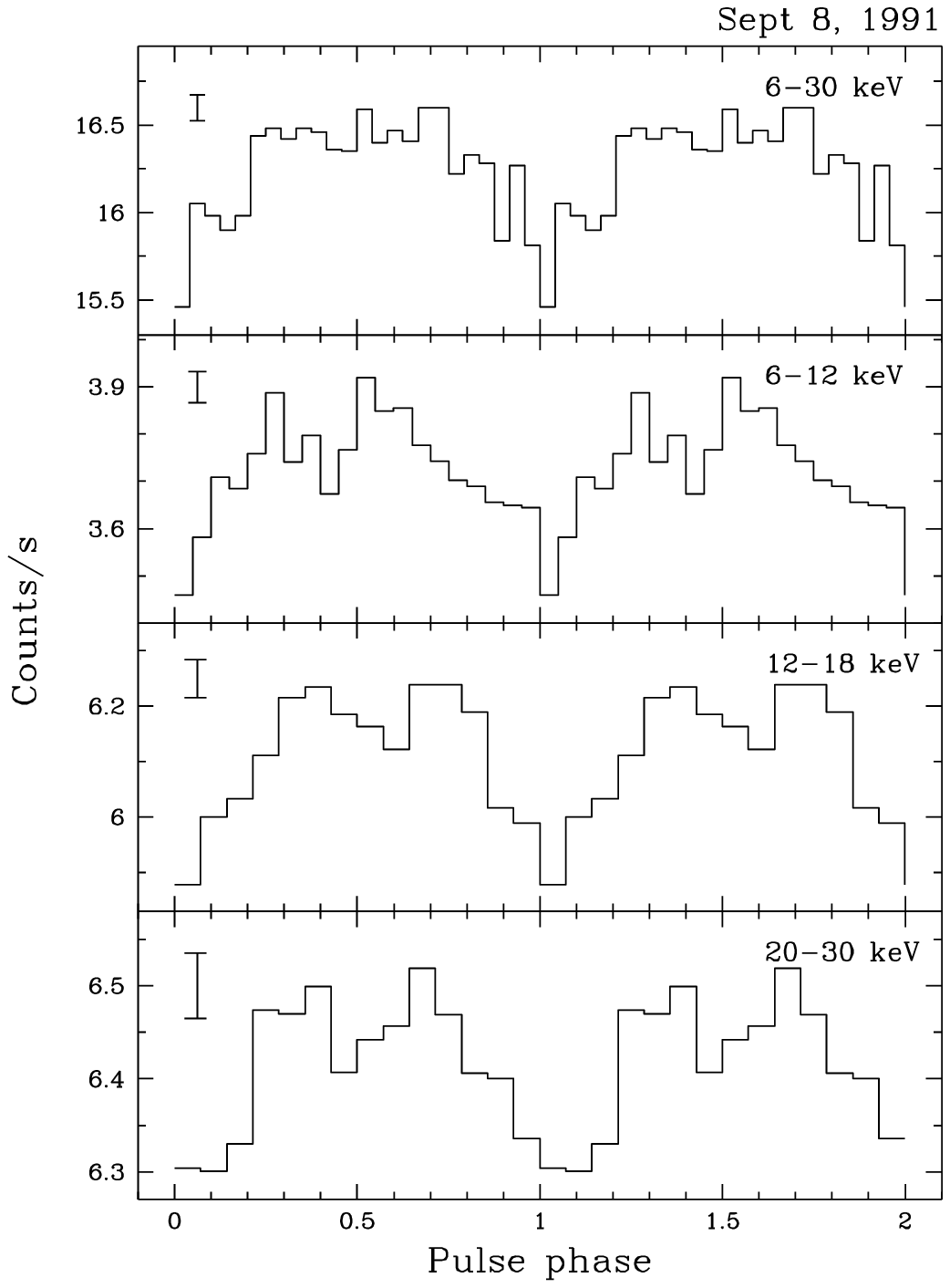}~b)
\caption{Observations of two different types of the
sourse pulse profile (single asymmetric peak and double peak) and their
dependence on the energy.} 
\label{LutovinovA_gx1_pp}
\end{vchfigure}

\subsection{GX1+4}

The hard X-ray pulsar GX1+4 in low-mass binary was observed several times
during the \Granat\ observatory operations in 1990-1992 (Lutovinov et
al. 1994).  During the observation on 23 Aug 1990 the source was detected in
the soft state where the observed photon spectrum was fitted well by a
simple powerlaw with a photon index $\alpha\sim2$. The X-ray luminosity in
the 3-20 keV range was equal to $\sim3.6\times10^{36}$ \ergs\ for a distance
of 8.5 kpc.

The X-ray pulsations with period of $114.66\pm0.01$ s were detected from the
source with a double peak structure in different energy bands on Oct 7, 1990
(Fig.\ref{LutovinovA_gx1_pp}a). During observations on Sept 7, 1991 the
pulse profile of GX1+4 showed a single broad feature with period
$116.14\pm0.13$ s (Fig.\ref{LutovinovA_gx1_pp}b). This change in the pulse
profile from a double to a single pulse structure in about one year
indicates either activation of the opposite pole of the neutron star if the
magnetic field is asymmetric or possibly a change in the beam pattern from a
pencil beam to a fan beam.

\subsection{GX5-1}

\begin{figure}[tb]
\begin{minipage}[t]{.45\textwidth}
\includegraphics[bb=100 325 460 680, width=\textwidth]{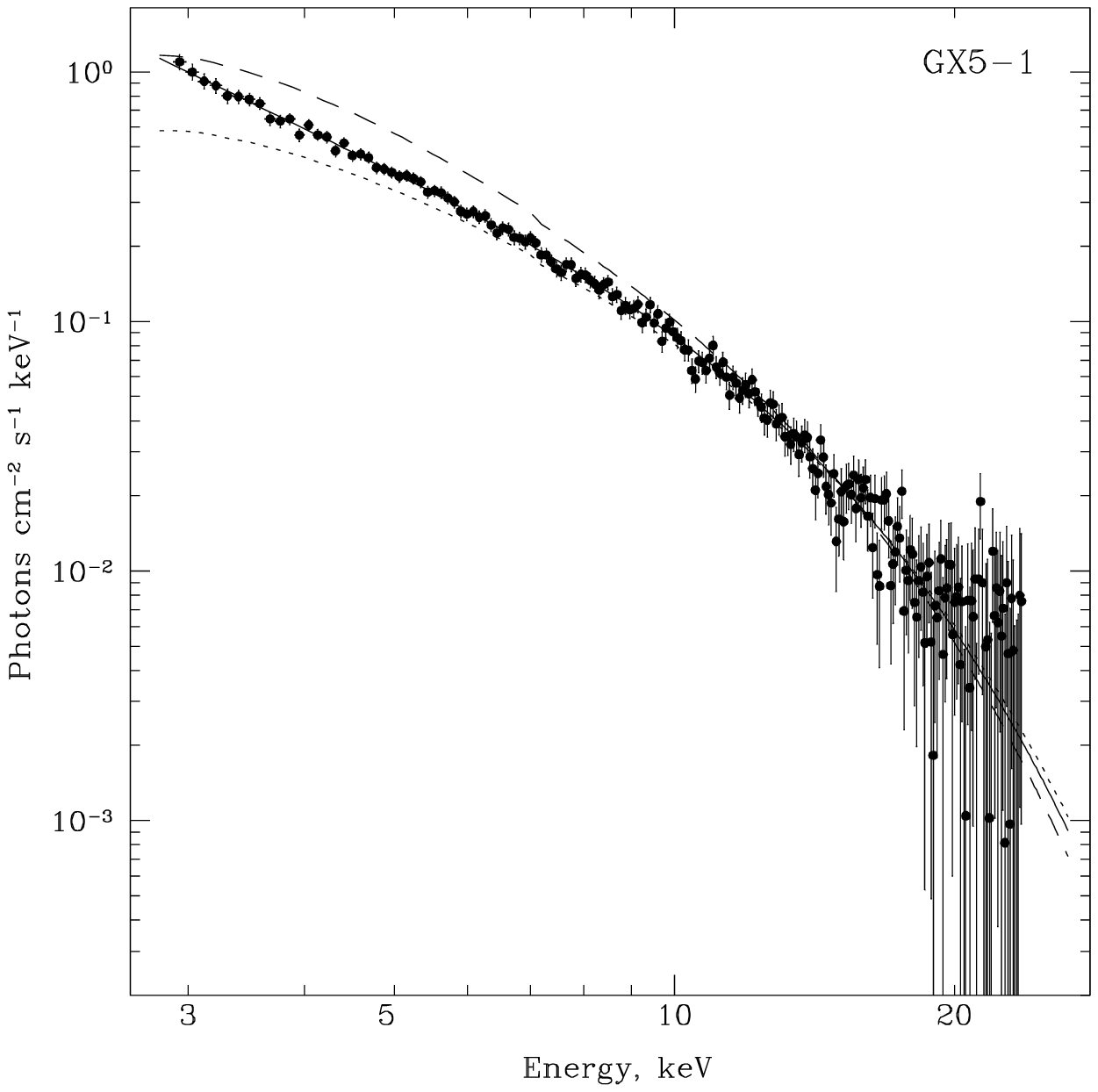}
\caption{The GX5-1 spectra variations as observed with ART-P in Sept-Oct
  1990. Lines are best-fit models for three different sessions of
  observations, points are the source spectrum obtained during one of them.}
\label{LutovinovA_gx5_sp}
\end{minipage}
\hfil
\begin{minipage}[t]{.45\textwidth}
\includegraphics[bb=100 325 460 680, width=\textwidth]{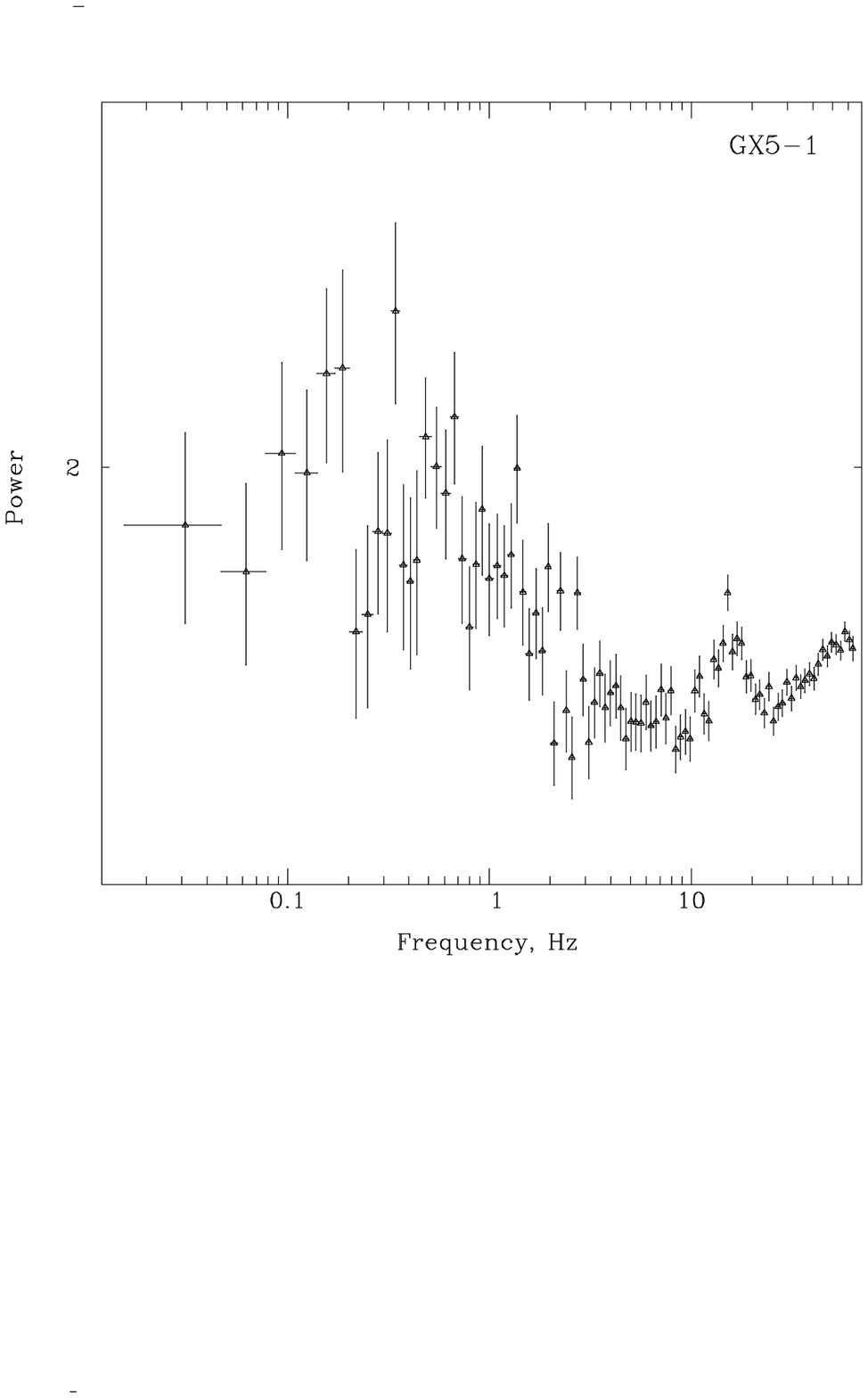}
\caption{The power spectrum (without the dead-time correction) of the ART-P
  light curve of GX5-1 obtained on Oct 20, 1990. The broad QPO feature in
  the range $13-20$ Hz is obviously visible. }
\label{LutovinovA_gx5_psp}
\end{minipage}
\end{figure}

GX5-1 is a second brightest source in X-rays after Sco X-1. It is called
Z-source after the approximate ``Z'' shape describe in an X-ray
colour-colour diagram and in an X-ray hardness-intensity diagram (Hasinger,
van der Klis 1989).  GX5-1 was observed by ART-P many times in 1990-1992
with intensity in the 3-20 keV energy band of $\sim1$ Crab. To investigate
the source spectral variations we applied to the spectra a few simple
models, such as Sunayev-Titarchuk Comptonization of soft photons on hot
electrons of plasma, optically thin thermal bremsstrahlung and Boltzmann
law. Figure \ref{LutovinovA_gx5_sp} illustrates this issue showing the
spectra measured on the autumn of 1990.

We estimated also the power spectra of the source by calculating the Fourier
amplitudes of the signal for the all sessions of observations. The QPO with
centroid frequency $\sim16$ Hz was detected on 20 Oct 1990
(Fig.\ref{LutovinovA_gx5_psp}). The measured frequency correspond to the
typical 'horizontal branch' QPO (van der Klis 1985).

\begin{acknowledgement}
A.L. thanks to the Organizing Committee and the Space Research Institute of
the Russian Academy of Sciences for the providing a financial support of the
participation in the Conference.
\end{acknowledgement}


\begin{thebibliography}{}

\bibitem{bib1} S. Grebenev et al., Adv. Space Res. \textbf{15}, (5)115
  (1995). 

\bibitem{bib2} S. Grebenev et al., in: Proceedings of the 2th INTEGRAL
  Workshop, ESA SP-382, p.183 (1996).   

\bibitem{bib1} P. den Hartog et al., A\&A  \textbf{400}, 633 (2003).

\bibitem{bib2} G. Hasinger,  M. van der Klis, A\&A  \textbf{225}, 79 (1989).

\bibitem{bib3} J. in't Zand, in: Proceedings of the 4th INTEGRAL Workshop,
  ESA SP-459, p.463 (2001).

\bibitem{bib4} E. Kuulkers et al., A\&A \textbf{382}, 503 (2002).

\bibitem{bib5} A. Lutovinov et al., Astron. Lett. \textbf{20}, 538 (1994).

\bibitem{bib6} A. Lutovinov et al., Astron. Lett. \textbf{27}, 501 (2001).

\bibitem{bib7} K. Makishima et al., Astrophys. J. \textbf{267}, 310 (1983).

\bibitem{bib8} S. Molkov et al., Astron. Lett. \& Comm. \textbf{38}, 141
  (1999). 

\bibitem{bib9} S. Molkov et al., A\&A \textbf{357}, L41 (2000).

\bibitem{bib10} S. Molkov et al., Astron. Lett. \textbf{27}, 363 (2001).

\bibitem{bib11} M. Pavlinsky et al., Sov. Astron. Lett. \textbf{18}, 88 (1992).

\bibitem{bib12} M. Pavlinsky et al., Astrophys. J. \textbf{425}, 110 (1994). 

\bibitem{bib13} M. Pavlinsky et al., Astron. Lett. \textbf{27}, 297 (2001).

\bibitem{bib14} L. Sidoly et al., Astrophys. J. \textbf{525}, 215 (1999).

\bibitem{bib15} D. Smith et al., Astrophys. J. \textbf{479}, L137 (1997).

\bibitem{bib12} R. Sunyaev et al., Sov. Astron. Lett. \textbf{16}, 59 (1990).

\bibitem{bib17} M. van der Klis et al., Nature \textbf{316}, 225 (1985).

\end{thebibliography}
\end{document}